# Basic non-linear effects in silicon radiation detector in detection of highly ionizing particles: registration of ultra rare events of superheavy nuclei in the long-term experiments.

Yu.S.Tsyganov

*FLNR , JINR, 141980 Dubna, Moscow reg., Russia*

Sources of non-linear response of PIPS detector, when detecting highly ionizing particles like recoils (EVR), fission fragments and heavy ions, including formation of large pulse-height defect (PHD) are considered. An analytical formula to calculate the recombination component of EVR's PHD is proposed on the base of surface recombination model with some empirical correction. PC-based simulation code for generating the spectrum of the measured recoil signal amplitudes of the heavy implanted nuclei is presented. The simulated spectra are compared with the experimental ones for the different facilities: the Dubna Gas Filled Recoil Separator (DGFRS), SHIP and RIKEN gas-filled separator. After the short reviewing of the detection system of the DGFRS, is considered the real-time matrix algorithm application aimed to the radical background suppression in the complete-fusion heavy-ion induced nuclear reactions. Typical examples of application in the long term experiments aimed to the synthesis of superheavy elements Z=112-118 are presented [1-9].

1**. Introduction**

Recently, more than 30 new nuclides with atomic numbers Z between 104 and 118 have been synthesized at the Dubna Gas-filled Recoil Separator (DGFRS) [1-7]. It should be noted that some of these experimental results have been clearly confirmed in independent experiments [7-9] involving the study of the chemical properties of the synthesized atoms. In order to succeed in detecting the synthesis of super heavy nuclides one has to pay attention to the following:

• an electromagnetic recoil separator design has to provide not only an acceptable value of the nuclide transportation efficiency (tens of percent), but also a significant suppression of the background products;

• the heavy-ion beam intensity has to be high enough to overcome the limited cross section for fusion followed by the evaporation of neutrons;

• a detection system has to provide a sufficient number of parameters in order to identify a nuclide. In additional, the design of the detection assembly has to provide for the suppression of the background products [10-12].

2. **Pulse amplitudes corresponding to evaporation residues (EVR) formed in heavy-ion induced complete-fusion nuclear reactions**

The multi-parameter events corresponding to production and decays of the super heavy elements (SHE) usually consist of the time-tagged recoil signal amplitudes and the alpha decay signal amplitudes. The amplitudes of the signals associated with one or two fission fragments (FF) arising in spontaneous fission (SF) might be present as well. The largest part of every detected multi-chain event consists of α-decays because of the following two reasons: 1) The possibility of measuring the energies of α-particles with a high accuracy provides sufficient information to define the atomic numbers of the original nuclides [13]. 2) The pulse amplitudes of EVRs and FF are observed with a significant pulse height defect value (PHD[1]); nevertheless, they are also of great interest since their presence at the beginning and end of each decay chain makes the identification process complete. F.P.Heßberger was the first who recognized the importance of such analysis and demonstrated its validity using the Monte Carlo simulation of SF decays of $^{256}$Rf nuclei implanted into a silicon radiation detector [15]. A simulation method for modeling the EVR spectra obtained from DGFRS is reported in [16, 17] in detail. A detailed description of a simulation approach for SF events can be found in [18, 19]. The successful application of these techniques to the data generated in an experiment which was carried out to investigate nuclides with atomic number Z=112 has been reported [14, 20]. The pulse amplitude of the EVR associated with one of three events in this particular case was anomalous.

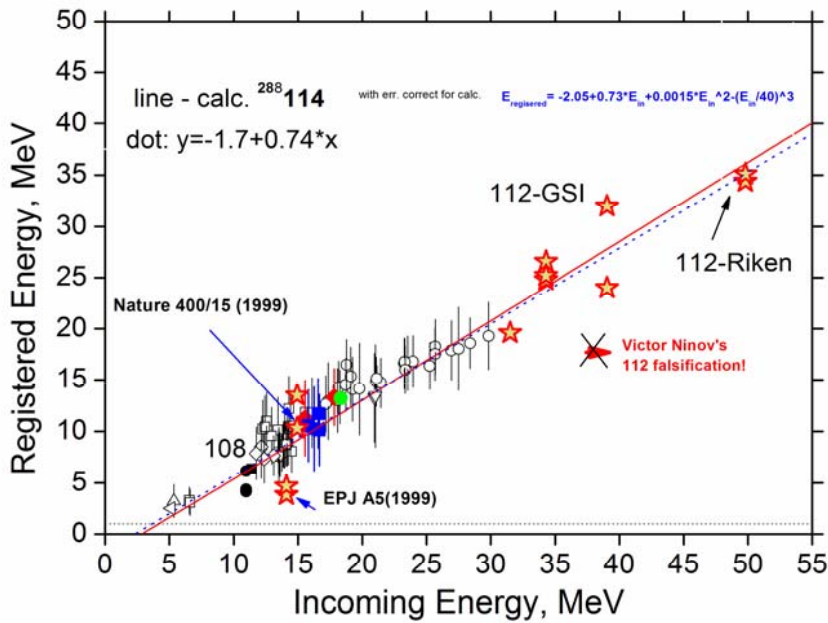

**Fig.1** The measured and calculated dependences of the values of the recoil energy signal amplitudes on actual incoming energies. For more details see the text.
$E_{Reg}(Z,A) = E_{Reg}(^{252}No) - k_1 E_{Inc}(A-252) - k_2 E_{Inc}(Z-102)$, $k_1 \approx 0.0169$; $k_2 \approx 0.058$. ($S_{Eff} \approx 10^3$, $<F> \leq 0.1$ V/μm)

---

[1] PHD = $\Delta_W + \Delta_R + \Delta_N$, dead layer, recombination and nuclear stopping components, respectively

The calculated and experimentally measured dependence of EVR energy signal amplitudes on the actual incoming energy for a DGFRS PIPS[2] detector is shown in Fig. 1. The solid curve is obtained by calculation [17]. The dotted line is an empirical calibration [18] measured with DGFRS for different heavy recoils. The experimental data obtained in SHIP, RIKEN [25] and VASILISSA [21] experiments are shown by stars. The experimental data corresponding to the nuclides with Z=112 measured in GSI [8,20] and RIKEN experiments are in good agreement with the calculation. We would like to note here that, in the case of the GSI experiments, the following two nuclear reactions were used: $^{238}$U+$^{48}$Ca and $^{68}$Ni+$^{206}$Pb. EVR amplitude which deviated significantly from the expectations of theory [14] was eliminated by the authors [20] after reanalysis. In order to produce the experimental data for $^{252}$No recoils, the nuclear reaction $^{206}$Pb+$^{48}$Ca→$^{252}$No+2n was used. For this case, a Mylar degrader was utilized, with thicknesses of 1 μm. The corresponding comparison between the calculated EVR energy spectra and the experimental data are presented in Fig.2a-b. The theoretical values demonstrate a good agreement with the measured ones, with the centroids deviating by 1.75 MeV which corresponds to 11%. Figure 2b shows the same comparison for the production of Z=112 nuclides in the reaction of $^{48}$Ca+$^{238}$U, where there is less experimental data; nevertheless, the deviation between experiment and simulation is only about 15% on the average.

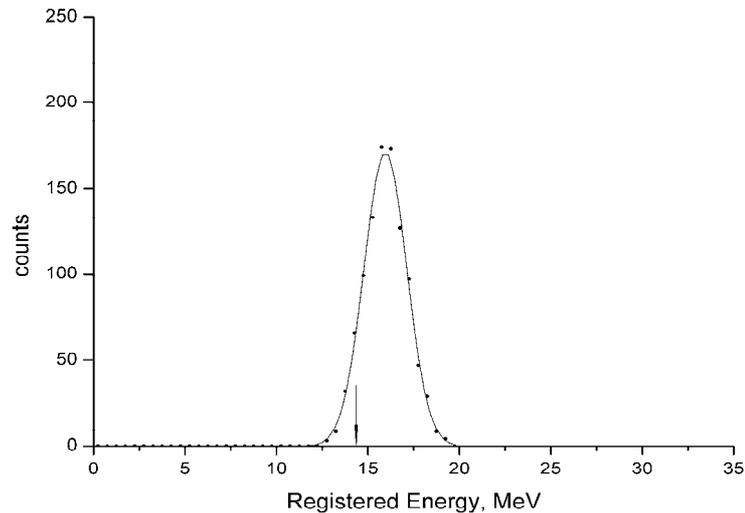

a)

---

[2] SHIP and Vasilissa detectors are manufactured by CANBERRA Semiconductors NV, Belgium.

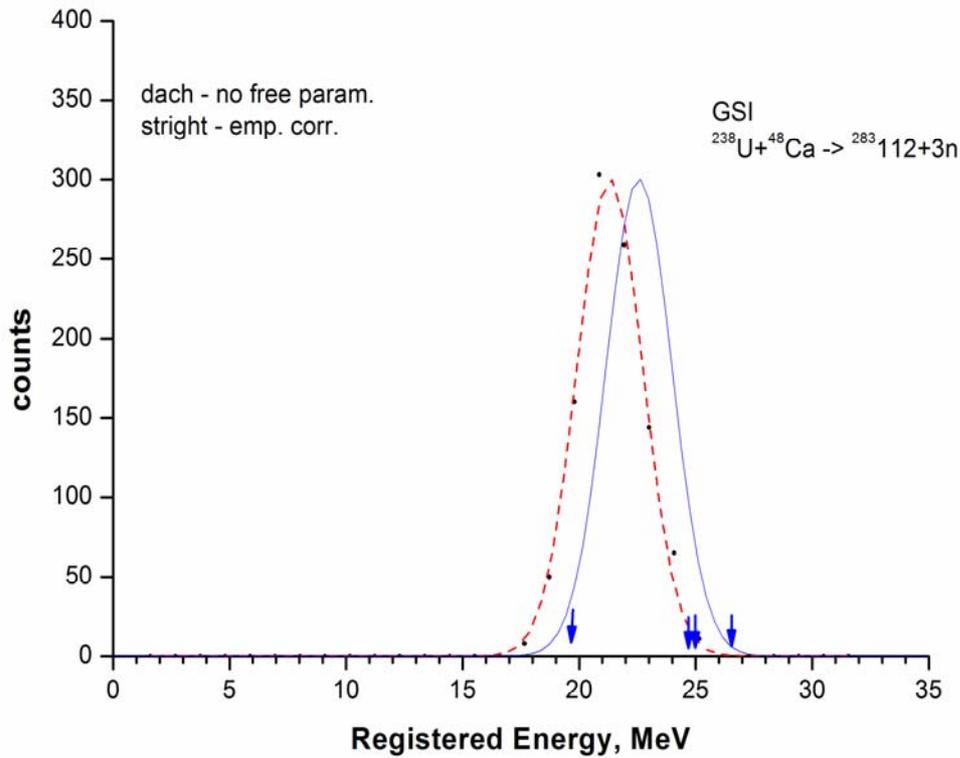

b)

**Fig.2** Simulated [present work] and experimental [8] EVR energy spectra for a) $^{252}$No nuclide, with 1 μm mylar degrader, $^{206}$Pb+$^{48}$Ca→$^{252}$No+2n. Centers of gravity are: 15.9 MeV (calc.) and 14.2 MeV (measured, see arrow); and b) Z=112 nuclide, no degrader, $^{238}$U+$^{48}$Ca → 112+3n. Calculated center of gravity is equal to 21.3 MeV. Since the experimental data in 2c is limited, no Gaussian is fit to these data. ( dot line - closed parameters of the simulation; line- with empirical corrections y=a+b$_1$x + b$_2$x$^2$; a=-1.22, b$_1$=0.3, b$_2$=-0.0062;) Relative recombination loss λ under simulation is equal to λ=gsT$_P$/R, where g-form-factor, s-effective recombination constant (~10$^3$ cm/s; g~0.5), R- particle range in silicon. Wilkins formula for stopping component of PHD is used.

3. **Recoil-FF signal amplitudes**

The present analysis for SHE nuclides does not always give a satisfactory agreement with measured values; for instance, in the synthesis of two recoil atoms with Z=114, see Ref. [21]. The experimental data obtained for an extended set of FF registered energy signals are shown in Fig. 3.

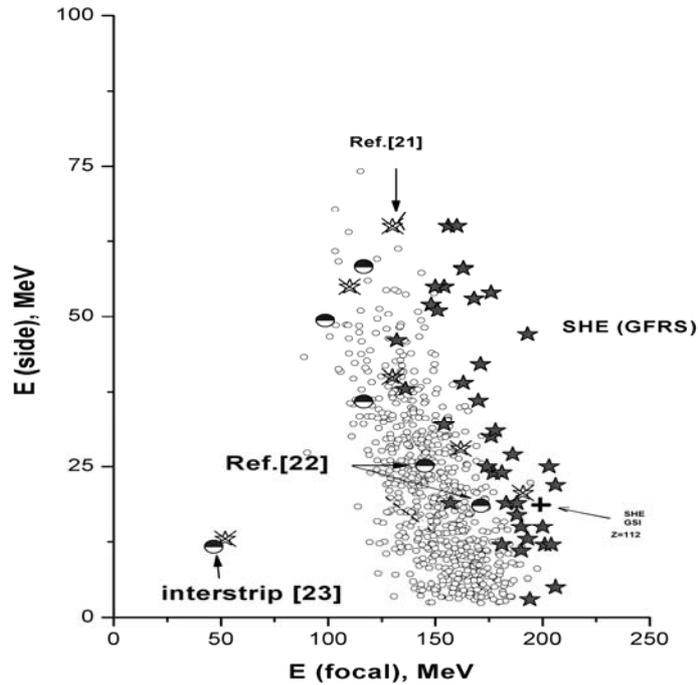

**Fig.3** Two-dimensional energy plot, SF events: the x axis corresponds to the energies of events detected by the main detector; the y axis corresponds to energies of the same events seen in the side detector. Open circles – $^{252}$No events; stars – SHE SF events measured with DGFRS; half-filled circles – data from a VASILISSA experiment with a SF energy calibration extrapolated from an α-calibration; open star with a cross on top of it – original data [21-23]; cross alone – Z=112 (GSI) [8].

The systematic shift of the groups of data points with respect to each other can be explained, in part, by systematic errors in the FF registered-energy scale which are caused by different experimental conditions. For this reason, let us consider a dimensionless energy value *k* equal to the ratio $E_{esc} / (E_{esc} + E_{main})$, where *esc* and *main* subscripts denote the energies detected by the side and main detectors, respectively. Use of the ratio diminishes the systematic shift of scales caused by different calibrations and can be introduced as a specific parameter for comparing the data. For complete-fusion reaction products, for which the implantation depth into the main silicon detector can be easily estimated, the *k*-parameter should decrease with an increase in the implantation depth. In Fig.4, the dependence of the *k*-parameter on implantation depth observed experimentally is compared with a simulation. It is seen that the values arising in the experiment described in [21] are well outside of the main trend obtained from theory[3]. In Fig.5, the calculated recoil spectrum corresponding to the experiment described in [21] is shown. The measured amplitudes for the two reported events are shown by arrows.

---

[3] Nuclear properties reported in [21] are not supported at this time by independent confirmation

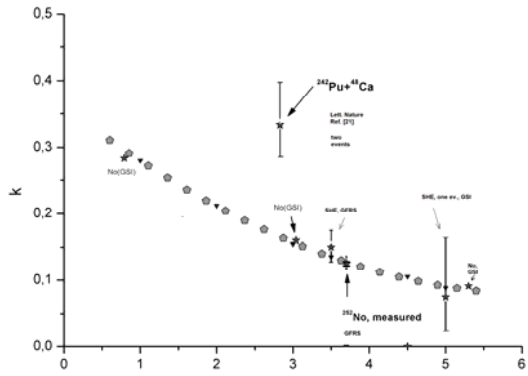

**Fig. 4** Dependence of the *k*-parameter against EVR implantation depth in Si (μm). Pentagons correspond to the results of a theoretical simulation. Stars denote the values extracted from measurements at different facilities, as indicated in the figure.

## 4. Detection of rare decays of SHE: "active correlations" technique

To detect ultra rare decays of SHE in the heavy-ion induced complete fusion nuclear reactions the "active correlations" technique has been designed and successfully applied during last six years. Namely with this technique it has became possible to provide a deep suppression of background products associated with the cyclotron. The idea of method is aimed at searching in real-time mode of energy-time-position recoil (EVR)-alpha links, using the discrete representation of the resistive layer of the position sensitive PIPS detector separately for events like "recoil" and 'alpha". Of course, preset parameters for "recoil" signals are calculated according to the reported above. More details of the present method are described in the Ref.'s [10-12]. PIPS detector and it's PC memory representation are shown in the Fig.5 a,b.

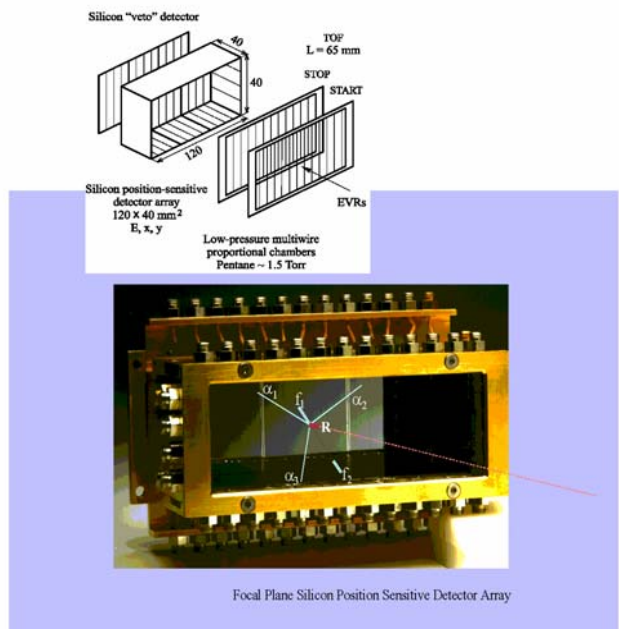

a)

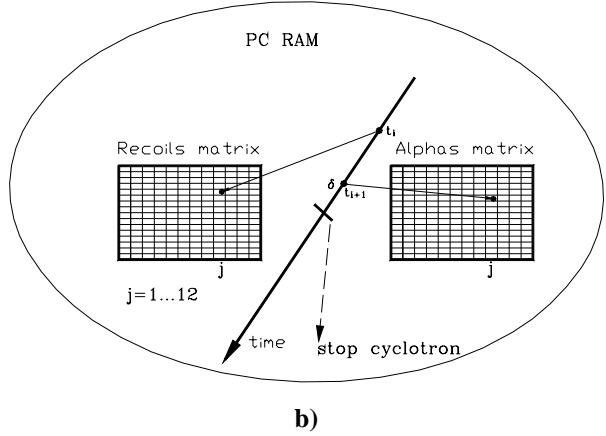

b)

**Fig.5**. PIPS focal plane detector of the DGFRS (a) and PC representation for EVR-alpha quick search (b). Detection module of the DGFRS consists of PIPS 12 strip position sensitive detector, eight side detectors fog higher geometric efficiency of particles detection, escaping the focal plane one, VETO detector to suppress long-path charged particles coming from cyclotron and creating no signal in TOF detector, and gaseous low pressure pentane filled TOF detector to detect EVR's and to suppress background particles. The second (vertical) matrix index sell number is specified as: $j = \text{int}\{N_{max} * [\frac{a_{iy}N_{yi}+b_{iy}}{a_i N_i + b_i} * (\frac{R0_i}{R_i}+1) - \frac{R0_i}{R_i} + \delta_i^{a,esc,EVR}]\}$, where $N_{max}$- maximum number (130-170), $R_i$- resistance value for a strip number #i (first index), $R0_i$ – additional (#i) resistor, $\delta$ – small corrections parameter which is different for some different particles(alpha-escaping alpha-EVR), **(a,b)** – set of calibration constants are obtained from test nuclear reactions, $N_{i, yi}$ – ADC's channels for main and coordinate scales, respectively.

In the **Fig.6** the result of application is shown for $^{249}$Am + $^{48}$Ca → 115 + 3n complete fusion reaction and in the Tab.1 the parameters of radical backgrounds suppression factor are shown in the column 2.

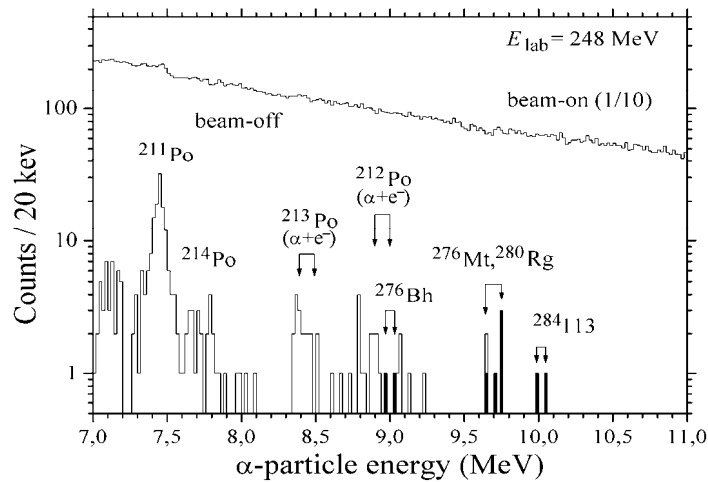

**Fig.6** Spectrum of alpha decays measured in the beam-OFF intervals

Table 1. Typical suppression factors when "active correlations" technique is applied.

| Reaction | An integral suppression factor (9-11 MeV) | Energy correlation interval (Eα, MeV) | Correlation time, s EVR - α | Beam pause, min |
|---|---|---|---|---|
| $^{238}$U+ $^{48}$Ca → 112 | 9,5 e+03 | 9,43 – 9,63/ 10,3-11,8 | 12/0,3 | 1 |
| $^{242}$Pu+$^{48}$Ca→114 | 4 e+03 | 9,9 – 10,35 | 4 | 1 |
| $^{245}$Cm+$^{48}$Ca→116 | 1,5 e+04 | 9,9 - 11 | 1 | 1 |
| $^{243}$Am+$^{48}$Ca→115 | 2.0 e+04 | 9,6 - 11 | 8 | 2 |
| $^{249}$Cf+$^{48}$Ca→118 | 1,1 e+04 | 9,9 - 12 | 1 | 1 |

## 5. Summary

1. based on:
   - theoretical models, EVR spectra simulations and empirical relations obtained from test reactions ,
   - Real-time matrix algorithm to search for pointer for potential forthcoming correlated sequence ,
   - DGFRS detection system,
   - U-400 cyclotron complex,

 a new radical technique of "active correlations" is designed, tested and successfully applied in the heavy ion-induced complete fusion nuclear reactions during last six years

- Detection of recoil-alpha correlated sequences in a real-time mode provides a deep suppression of beam associated backgrounds, when ultra rare alpha decays are detected. It provides more clear event detection and identification in long-term experiments aimed to the synthesis of SHE

- A loss in the value of a total experiment efficiency is negligible, whereas an additional integral background suppression factor of about $10^4$ in the vicinity of 10 MeV energy interval has been achieved

2. Model calculations for EVR and SF events spectra are performed and compared with the different experimental data. This provides a framework for the criticism of the interpretation of the experimental results in the work [21]. This criticism is supported by the results of independent experiments, basing on the measurement of Z=112 nuclide decay properties (see also [26])

3. For the case of lower PIPS detector over-depletion ratio, the formula for calculation of EVR registered energy signal amplitude in for the incoming energy interval of 5 - 50 MeV is proposed with the Z,A-depending correction coefficients (Fig.1 and it's caption).

4. A dimension-less energy parameter $k(r) = E_{esc}/(E_{main}+E_{esc})$ is proposed for critical amplitude analysis of spontaneous fission decays of implanted nuclei for the case of both fragments are detected by focal plane and side detectors.

( r – calculated implantation depths in silicon)

**Acknowledgements**

This work has been carried out with the support of Grant № 07-02-0029 from the Russian Foundation for Basic Research. The author would like to thank Drs. V.K.Utyonkov and K.Moody[4], for supporting the present work.

---

[4] LLNL, Livermore, Univ. of California, USA